\documentclass{article}
\usepackage{spconf,amsmath,graphicx}

\usepackage{enumitem}
\setlist{nosep, leftmargin=14pt}

\usepackage{mwe} 

\usepackage{amssymb}
\usepackage{latexsym}

\usepackage{url}
\usepackage[dvipsnames]{xcolor}
\def\rm{\color{Mahogany}}

\usepackage{hyperref}

\usepackage{subcaption} 
\usepackage{amsmath, amsfonts,graphics} 
\usepackage{graphicx}

\usepackage{tabularx}

\usepackage{svg}

\usepackage{soul}

\usepackage{wrapfig}

\DeclareMathOperator*{\argmin}{arg\,min}

\title{Joint Deep Learning For Improved Myocardial Scar Detection From Cardiac MRI}
\address{Author Affiliation(s)}
\name{
\parbox{\linewidth}{
\centering
Jiarui Xing$^{ a}$ \qquad
Shuo Wang$^{ c}$ \qquad \\
Kenneth C. Bilchick$^{ c}$ \qquad
Amit R. Patel$^{ c}$ \qquad 
Miaomiao Zhang$^{ a, b}$
}}

\address{
\parbox{\linewidth}{
\centering
$^{a}$ Department of Electrical and Computer Engineering, University of Virginia, USA \\
$^{b}$ Department of Computer Science, University of Virginia, USA  \\
$^{c}$ School of Medicine, University of Virginia Health System, USA
}}

\begin{document}
%
\maketitle
%
%
\begin{abstract}
Automated identification of myocardial scar from late gadolinium enhancement cardiac magnetic resonance images (LGE-CMR) is limited by image noise and artifacts such as those related to motion and partial volume effect. This paper presents a novel joint deep learning (JDL) framework that improves such tasks by utilizing simultaneously learned myocardium segmentations to eliminate negative effects from non-region-of-interest areas. In contrast to previous approaches treating scar detection and myocardium segmentation as separate or parallel tasks, our proposed method introduces a message passing module where the information of myocardium segmentation is directly passed to guide scar detectors. This newly designed network will efficiently exploit joint information from the two related tasks and use all available sources of myocardium segmentation to benefit scar identification. We demonstrate the effectiveness of JDL on LGE-CMR images for automated left ventricular (LV) scar detection, with great potential to improve risk prediction in patients with both ischemic and non-ischemic heart disease and to improve response rates to cardiac resynchronization therapy (CRT) for heart failure patients. 
Experimental results show that our proposed approach outperforms multiple state-of-the-art methods, including commonly used two-step segmentation-classification networks, and multitask learning schemes where subtasks are indirectly interacted.


\end{abstract}

\section{Introduction}

Myocardial scarring is an important factor that leads to cardiac dysfunction and poor prognosis of patients with heart diseases~\cite{krittayaphong2011prevalence,turkbey2015prevalence}. Identifying the presence and extent of myocardial scars is important to many clinical interventions. For instance, it can significantly increase the response rate of cardiac resynchronization therapy (CRT), since contractile improvements are often shown on scar-free sites~\cite{ypenburg2007impact}. Another example is that the quantification of scar tissue can provide prognostic information about the risk for arrhythmia in patients with nonischemic cardiomyopathy undergoing cardioverter-defibrillator implantation~\cite{neilan2013cmr}. With rapid development of advanced imaging techniques, LGE-CMR has become one of the most promising and noninvasive methods to determine myocardial scar burden and guide therapeutic decisions~\cite{kim1999relationship}. However, the variable quality of LGE images due to image noises, motion artifacts, and partial volume effect makes the task of automated scar detection challenging.

Inspired by the great performance of deep learning methods, recent research has made the first attempt to address this issue by employing deep neural networks to segment scar directly from LGE-CMR images~\cite{moccia2018automated,moccia2019development}. Despite relatively low Dice scores, these works demonstrated that the predicted scar segmentations can be substantially improved with provided myocardium segmentation labels. Later, two-step approaches were developed to achieve better performance by cascading the task of myocardium and scar segmentation sequentially in two separate networks~\cite{zabihollahy2020fully}. While the scar segmentation accuracy was improved, the performance of two-step methods heavily depends on the quality of myocardium segmentation. Another mainstream of incorporating myocardium segmentations is through multitask learning (MTL). In contrast to two-step approaches, MTL performs both tasks simultaneously and allows mutual benefit from each other~\cite{xu2018mutgan,yang2020simultaneous}. However, MTL methods are fully data-driven and the learned relationship between tasks is unclear. With little to no knowledge on how the task relationship is utilized, the performance of MTL can be unstable even with fine-tuned parameters. In addition, MTL heavily depends on the quality of training data. It may learn an inaccurate task relationship that misleads the learning towards suboptimal models when the data is low quality with high noises.


In this paper, we propose a novel joint deep learning (JDL) framework that efficiently solicits the knowledge of spatial relationships between the myocardium and myocardial scars from jointly learned networks. In contrast to previous methods treating related tasks in a separate or parallel manner, our model (i) explicitly enforces a mutual guidance between the myocardium and scar segmentation tasks in training, and (ii) directly incorporates the prior information provided by predicted myocardium segmentations to identify the scars. We develop an interactive joint deep network that connects myocardium segmentation and myocardial scar detection by a message-passing module. More specifically, we use the predicted myocardium probability maps as weighting masks on the LGE images to eliminate the negative effects of non-ROI regions in scar detection. Meanwhile, the myocardium segmentation network receives extra supervision from the scar maps, which captures the scar region more accurately.


To the best of our knowledge, we are the first to develop JDL models for myocardial scar detection with learned myocardium segmentations from LGE-CMR images. Experimental results show that our model outperforms a direct segmentation network on original LGE images~\cite{moccia2019development}, fully automatic two-step approaches~\cite{zabihollahy2020fully}, and recent MTL schemes~\cite{yang2020simultaneous}. The outcome of this research has great potential to impact various image-guided clinical systems, such as locating optimal left-ventricular-lead sites that are scar-free for CRT~\cite{wong2013influence}, providing more accurate prognostic information in patients with known or suspected coronary artery disease~\cite{krittayaphong2011prevalence}, and making the best possible therapeutic decisions for coronary revascularization~\cite{sharma2021modified}.

\section{A Joint Deep Network For Scar Detection}
Consider a set of training images $I_1, \cdots, I_N$ with ground truth labels of myocardium and myocardial scars, where $N$ is the number of images. Our JDL framework simultaneously learns the tasks of myocardium and scar segmentation in an interactive manner. With a joint optimization scheme, our model produces two different outputs: a myocardium segmentation probability map and a binary scar map. The myocardium probability map is used as a ROI-mask that is iteratively passed to the learning process of scar detection. Next, we will introduce the network architecture and optimization schemes.  

\subsection{Joint Network Architecture}
\begin{figure}[h]
    \centering
    \includegraphics[width=\linewidth]{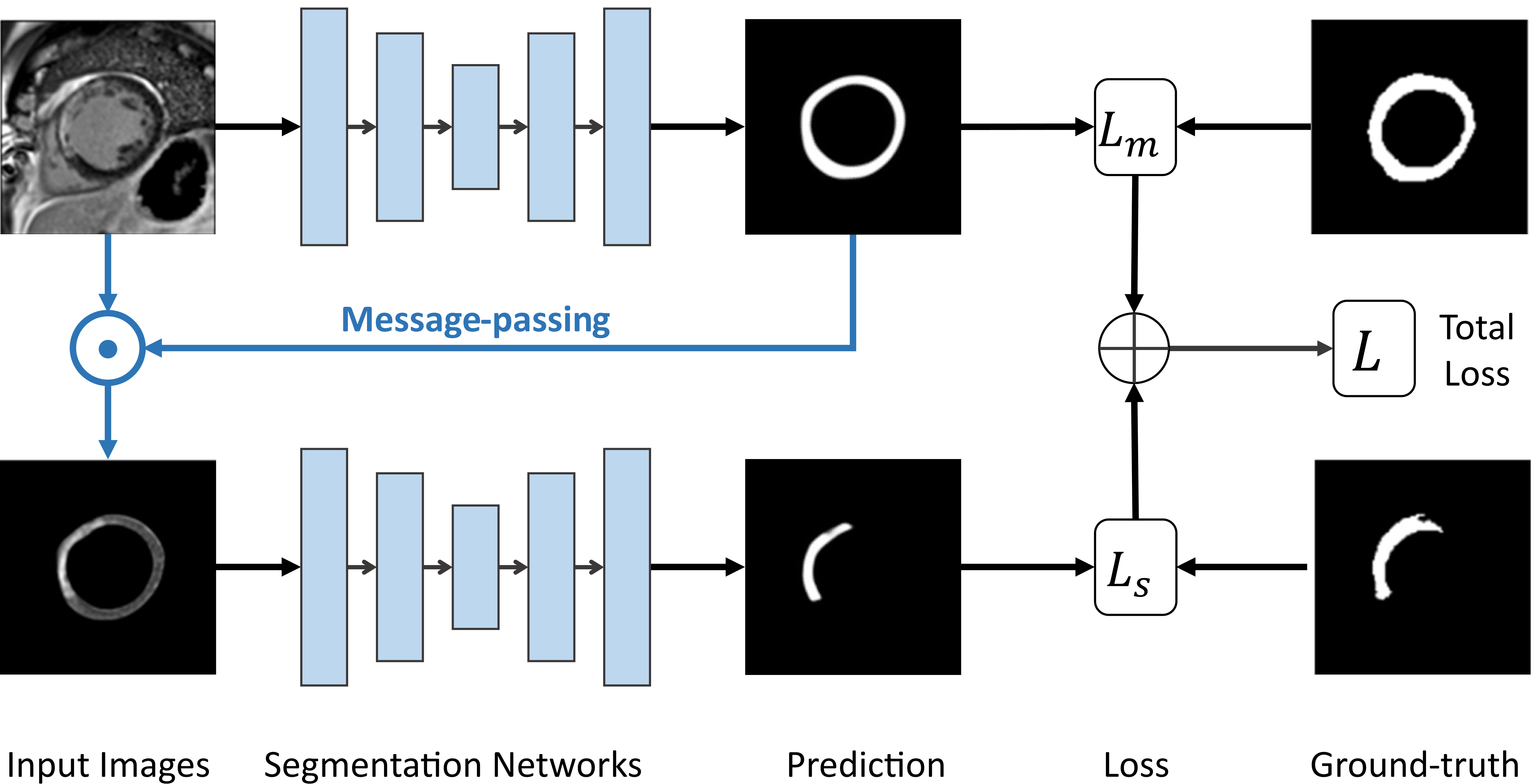}
    \caption{The overview of our proposed framework, where the myocardium segmentation network communicates with the scar segmentation network via the message-passing module, and the two networks are jointly trained. The $\bigodot$ means element-wise multiplication and refers to masking operation, and $\bigoplus$ means summation for the joint optimization.}
    \label{fig:pipeline}
\end{figure}

Our designed network is conceptually simple and modifies two-step approaches~\cite{zabihollahy2020fully} to jointly optimized networks through a message passing module. The first myocardium segmentation network takes LGE-CMR as an input and generates a probability map. A message passing module then combines the information of probability maps with the original images to reproduce a masked input data for the scar segmentation network. This effectively suppresses noises or artifacts from non-ROI regions in scar detection. It is worthy to mention that our framework is highly flexible to various types of segmentation networks for both tasks. In this paper, we employ pre-trained TransUNet~\cite{ridnik2021imagenet21k}, which shows superior performance on medical image segmentation tasks over other commonly used networks such as U-Net~\cite{ronneberger2015u} An overview of our proposed approach is presented in Fig.~\ref{fig:pipeline}.

\subsection{Loss and Optimization}
Consider a set of ground truth segmentation labels $Y$, we use a weighted sum of cross entropy and Dice loss $L(\cdot, \cdot)$ for both segmentation networks, i.e.,
$$L(Y, \hat{Y}(\theta))=-\sum Y \log \hat{Y}(\theta) - w \sum  (1-\frac{2Y\hat{Y}(\theta) + \epsilon}{Y+\hat{Y}(\theta)+\epsilon}),$$
where $\hat{Y}(\theta)$ represents network prediction with parameter $\theta$, $w$ is a weighting parameter, and $\epsilon$ is a small smoothing constant. In this paper, we set $\epsilon=1e-6$ in all our experiments.

Supervised by manually delineated myocardium $M$ and myocardial scars $S$, our model jointly learns to predict myocardium segmentation probability maps $\hat{M}(\theta_M)$ and binary scar maps $\hat{S}(\theta_S)$. A message-passing module effectively utilizes the predicted myocardium segmentations to mask out non-ROI regions from original images, which provides new input data to the scar network. The total loss function is then defined by a weighted sum of the regularized myocardium segmentation loss $L_m$ and the scar segmentation loss $L_s$
\begin{align}
\label{eq:loss}
    L(\theta_M, \theta_S) = & \lambda \sum_{i=1}^N L_m \left(M_i, \hat{M}_i(\theta_M); I_i \right)\nonumber\\
    & + L_s \left(S_i, \hat{S}_i(\theta_S); \hat{M}_i(\theta_M) \odot I_i \right)  \nonumber \\ 
    & +\beta_M \text{Reg}(\theta_M) + \beta_S \text{Reg}(\theta_S),
\end{align}
where $\{\lambda, \beta_M, \beta_S\}$ are weighting parameters, $\odot$ denotes an element-wise production), and $\text{Reg}(\cdot)$ represents network regularization, which is set to $L_2$ weight decay in our experiments.

\textbf{Joint network optimization.} We use a joint optimization scheme to minimize the loss in Eq.~\eqref{eq:loss}. Our scar segmentation network is trained using the output of the myocardium segmentation network, whereas the myocardium segmentation network is trained by taking a sum of weighted gradients from both loss functions $L_m$ and $L_s$. The optimization of Eq.~\eqref{eq:loss} splits into two parts:     
\begin{align}
    \hat{\theta}_M = &\argmin_{\theta_M}  \lambda \sum_{i=1}^N L_m \left(M_i, \hat{M}_i(\theta_M); I_i \right) \nonumber\\& +L_s \left(S_i, \hat{S}_i(\theta_S); \hat{M}_i(\theta_M) \odot I_i \right) \nonumber \\
    & +\beta_M \text{Reg}(\theta_M), \nonumber \\
    \hat{\theta}_S = &\argmin_{\theta_S} L_s \left(S_i, \hat{S}_i(\theta_S); \hat{M}_i(\theta_M) \odot I_i \right) + \beta_S \text{Reg}(\theta_S). \nonumber 
\end{align}
This shows that our model explicitly allows two tasks mutually guided through the message-passing module. In practice, we train the two subnetworks simultaneously. A detailed description of parameter settings will be introduced in the experiment section.

\section{Experimental Results}
\subsection{Experimental Settings} 

We test the proposed JDL framework on a newly collected dataset of LGE-CMR images in our institute. To validate the performance of our model, we compare it with three types of baseline algorithms: a direct scar segmentation network on original image data, a two-step myocardium and scar segmentation network, and a multitask learning network with hard-parameter sharing. For fair comparison, the backbone network of all methods is TransUNet~\cite{chen2021transunet} with the transformer part pre-trained on imageNet-21K~\cite{ridnik2021imagenet21k}. To evaluate volume overlap between the predicted segmentation A and the manual segmentation B, we compute the Dice similarity coefficient $DSC(A, B) = 2\left(|A| \cap |B|\right)/\left(|A| + |B|\right)$, where $\cap$ denotes an intersection of two regions. 

\textbf{Data.} We have been collecting LGE-CMR images with the size of $224\times224$ from $168$ heart failure patients who had CMR studies prior to undergoing the treatment of CRT. All ground truth labels, including segmentation of myocardium and scar maps involved in our experiment were manually delineated by experienced medical doctors. $510$ LGE-CMR images from $140$ patients are used for training, and $125$ images from $28$ different patients are used for testing. All images are normalized to $[0,1]$ before sending to the network. We generated random affine transforms, including scaling, rotating, translating and shearing, for data augmentation. While the data used in this paper is not publicly available online, upon the acceptance of this manuscript, we commit to share all images along with myocardium and scar labels if an institutional data use agreement is in place.

\textbf{Network training and hyper-parameter tuning.} The joint network is trained on GPU units with Adam optimizer~\cite{DBLP:journals/corr/KingmaB14}. We set the batch size and the number of epochs to $40$ and $500$ respectively. We have the learning rate $\eta$, drop-out rate $p$, loss weight $\lambda$ and $L_2$ weight decay weights $\beta_M$ and $\beta_S$ as hyper-parameters to be tuned using Bayesian optimization with Gaussian processes implemented by wandb~\cite{wandb} package. The empirical optimal hyper-parameters for JDL are $\hat{\eta} = 4.73e-4$, $p=0.39$, $\hat{\lambda}=1.02$ and $\beta_M=\beta_S=5.58e-6$.

\textbf{Compared methods.} We compared our proposed approach with 3 other approaches: the direct segmentation approach, the two-step approach and the multi-task learning approach. In the direct segmentation approach a segmentation network is directly trained to predicted the scar segmentation mask given one LGE MR image. In the two-step approach, we train a myocardium segmentation network, and separately train a scar segmentation network that takes MR image masked by manual labeled myocardium segmentation mask. In the testing stage, we first predict the myocardium mask using the myocardium segmentation network and use them to mask the LGE MR images. Then, the masked images are sent to the scar segmentation network to generate scar segmentation mask. All the networks mentioned above use the TransUNet~\cite{ridnik2021imagenet21k} as backbone. For the multi-task learning approach, we adopted the multi-head network structure, where the network has on shared sub-network followed by two sub-branch networks, one of which generates myocardium segmentation mask and the other generates scar segmentation mask. In the implementation, we duplicated the decoder (i.e. the upscaling convolutional layers) of the TransUNet~\cite{ridnik2021imagenet21k} as two sub-branch networks. The loss function of (sub-branch) networks is the same with our method as Eq.\ref{eq:loss}.


\subsection{Results \& Discussion} 

\begin{figure}[!htb]
    \centering
    \includegraphics[width=\linewidth]{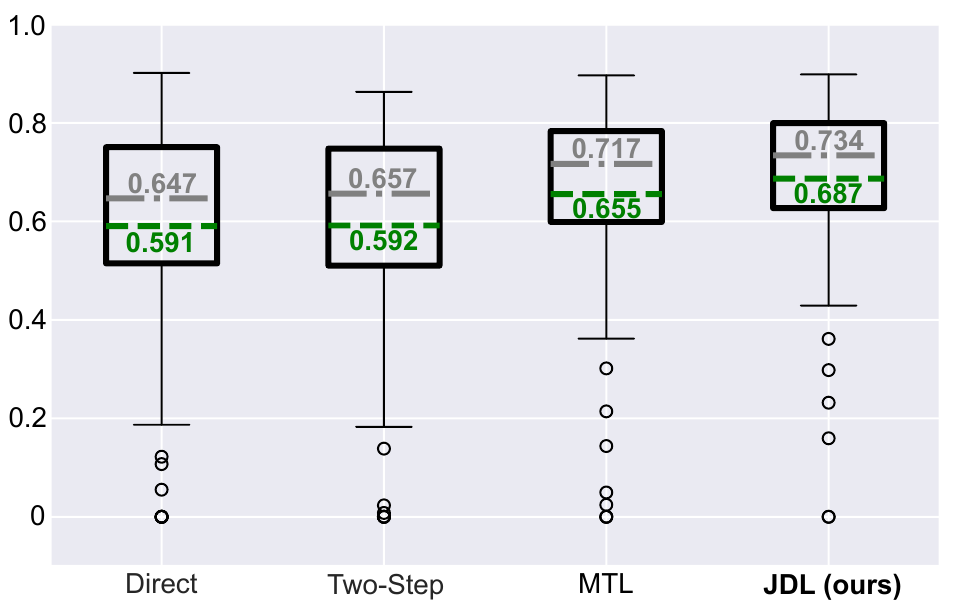}
    \caption{Scar  Segmentation Dice score of Different Methods}
    \label{fig:boxplot-scar}
\end{figure}


\begin{figure*}[!htb]
    \centering
    \includegraphics[width=\linewidth]{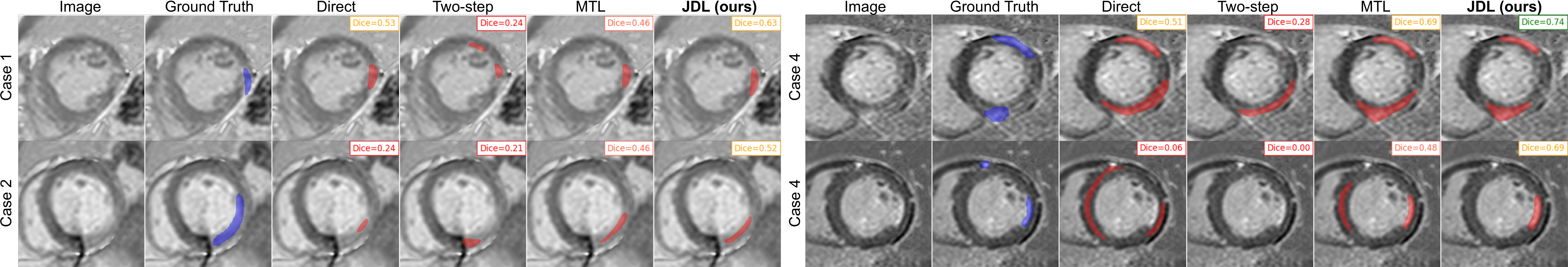}
    \caption{A comparison of predicted scar segmentations between our method and other approaches. Left to right: input images, manually delineated ground truth labels, direct segmentation on original images, two-step approach, multitask learning (MTL), and our joint deep learning (JDL) framework. }
    \label{fig:examples-scar}
\end{figure*}

The comparison of Dice scores of all methods are reported in Fig.~\ref{fig:boxplot-scar}. It shows that our proposed JDL framework outperforms all other approaches in terms of both accuracy and robustness. Examples of segmentation results are shown in Fig.~\ref{fig:examples-scar}, which also demonstrates that our method achieves the closest scar segmentations to manual delineations  (ground truth) provided by experts.

\section{Discussion}
\subsection{Interpretation of JDL Framwork}
\begin{figure}[!h]
    \centering
    \includegraphics[width=\linewidth]{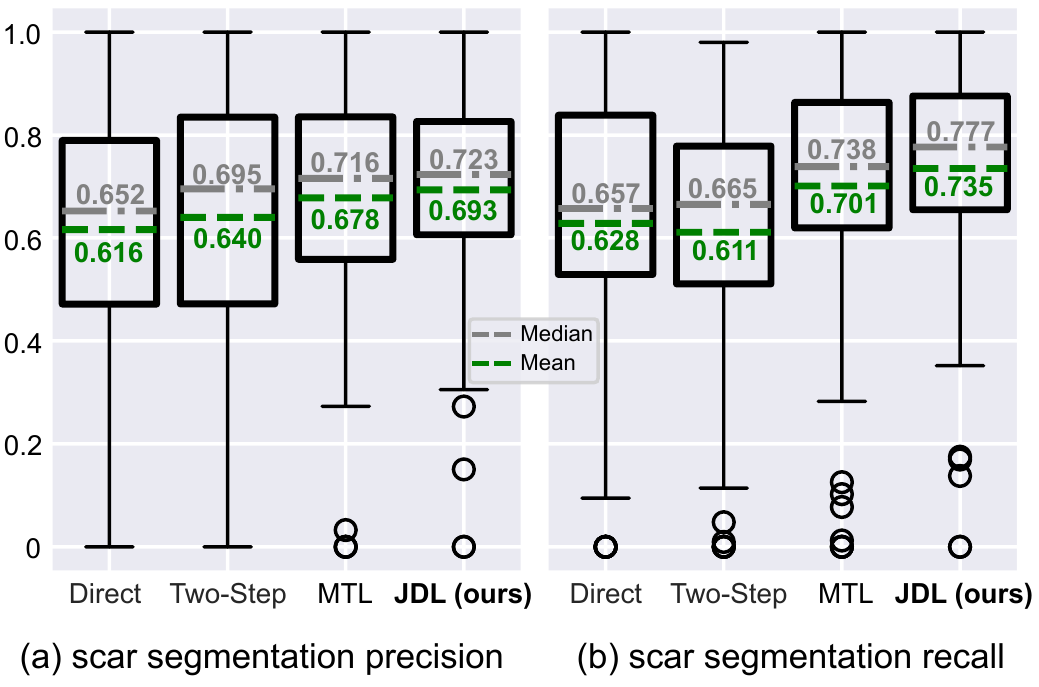}
    \caption{Scar segmentation precision and recall of all methods}
    \label{fig:precision-and-recall}
\end{figure}

To further understand why our JDL achieves better performance, we also computed the precision and recall of the segmentation results. For each segmentation result, they are computed as $\text{Precision}(S, \hat{S}) = (S\cap \hat{S}) / \|  \hat{S} \|$ and $\text{Recall}(S, \hat{S}) = (S\cap \hat{S}) / \| S \|$
where $S$ and $\hat{s}$ are the ground-truth and predicted masks, respectively. The segmentation model with high precision tends to capture all possible areas and misses less, while the model with high recall tends to avoid false positive prediction and tries to ensure all the captured regions are indeed the target region. The scar segmentation precision and recall of different methods are shown in Fig.~\ref{fig:precision-and-recall}. While our JDL model has both the best precision and recall performance, its recall is significantly higher than all other methods, which indicates that our JDL model is less likely to wrongly report the existence of scar at a scar-free region, and this also matches the examples given in Fig.\ref{fig:examples-scar}. 

\subsection{Effect on Myocardium Segmentation Task}
While in general, when multiple tasks are trained jointly, the performance improvement of all tasks is expected. However, the mean Dice scores of the myocardium segmentation of the two-step approach, the MTL approach and our approach are $0.894$, $0.900$ and $0.891$, respectively. In other words, the joint training strategy has little effect on the myocardium segmentation task, while it does improve the scar segmentation task. This raise further interpretability question to those approaches, which is crucial to build trustable algoroithms and could be the future work of the proposed JDL method.



\section{Conclusion}
In this paper, we proposed a novel joint deep neural network for an improved performance of myocardial scar detection with learned myocardium segmentations. In contrast to existing deep learning based approaches that train these tasks in a separate or parallel manner, our methods incorporates a message-passing module that effectively passes information within the subnetworks. Therefore, our model is able to explicitly use the information of learned myocardium segmentations to guide the scar detection. Meanwhile, the myocardium segmentation network receives additional supervision from the scar labels, which improves the predicted segmentation accuracy. Experimental results show that our method achieved the state-of-the-art performance on myocardial scar segmentation. Our work has great potential to impact various image-guided clinical systems, such as identifying optimal pacing sites that are scar-free for efficient treatment of CRT~\cite{wong2013influence}. 
We commit to make code produced in this work publicly available online upon the acceptance of this manuscript. 
\bibliographystyle{IEEEbib}
\bibliography{refs}

\begin{thebibliography}{10}

\bibitem{krittayaphong2011prevalence}
Rungroj Krittayaphong, Pairash Saiviroonporn, Thananya Boonyasirinant, and
  Suthipol Udompunturak,
\newblock ``Prevalence and prognosis of myocardial scar in patients with known
  or suspected coronary artery disease and normal wall motion,''
\newblock {\em Journal of Cardiovascular Magnetic Resonance}, vol. 13, no. 1,
  pp. 1--8, 2011.

\bibitem{turkbey2015prevalence}
Evrim~B Turkbey, Marcelo~S Nacif, Mengye Guo, Robyn~L McClelland, Patricia~BRP
  Teixeira, Diane~E Bild, R~Graham Barr, Steven Shea, Wendy Post, Gregory
  Burke, et~al.,
\newblock ``Prevalence and correlates of myocardial scar in a us cohort,''
\newblock {\em Jama}, vol. 314, no. 18, pp. 1945--1954, 2015.

\bibitem{ypenburg2007impact}
Claudia Ypenburg, Martin~J Schalij, Gabe~B Bleeker, Paul Steendijk, Eric
  Boersma, Petra Dibbets-Schneider, Marcel~PM Stokkel, Ernst~E van~der Wall,
  and Jeroen~J Bax,
\newblock ``Impact of viability and scar tissue on response to cardiac
  resynchronization therapy in ischaemic heart failure patients,''
\newblock {\em European heart journal}, vol. 28, no. 1, pp. 33--41, 2007.

\bibitem{neilan2013cmr}
Tomas~G Neilan, Otavio~R Coelho-Filho, Stephan~B Danik, Ravi~V Shah, John~A
  Dodson, Daniel~J Verdini, Michifumi Tokuda, Caroline~A Daly, Usha~B Tedrow,
  William~G Stevenson, et~al.,
\newblock ``Cmr quantification of myocardial scar provides additive prognostic
  information in nonischemic cardiomyopathy,''
\newblock {\em JACC: Cardiovascular Imaging}, vol. 6, no. 9, pp. 944--954,
  2013.

\bibitem{kim1999relationship}
Raymond~J Kim, David~S Fieno, Todd~B Parrish, Kathleen Harris, Enn-Ling Chen,
  Orlando Simonetti, Jeffrey Bundy, J~Paul Finn, Francis~J Klocke, and Robert~M
  Judd,
\newblock ``Relationship of mri delayed contrast enhancement to irreversible
  injury, infarct age, and contractile function,''
\newblock {\em Circulation}, vol. 100, no. 19, pp. 1992--2002, 1999.

\bibitem{moccia2018automated}
Sara Moccia, Riccardo Banali, Chiara Martini, Giuseppe Moscogiuri, Gianluca
  Pontone, Mauro Pepi, and Enrico~Gianluca Caiani,
\newblock ``Automated scar segmentation from cmr-lge images using a deep
  learning approach,''
\newblock in {\em 2018 Computing in Cardiology Conference (CinC)}. IEEE, 2018,
  vol.~45, pp. 1--4.

\bibitem{moccia2019development}
Sara Moccia, Riccardo Banali, Chiara Martini, Giuseppe Muscogiuri, Gianluca
  Pontone, Mauro Pepi, and Enrico~Gianluca Caiani,
\newblock ``Development and testing of a deep learning-based strategy for scar
  segmentation on cmr-lge images,''
\newblock {\em Magnetic Resonance Materials in Physics, Biology and Medicine},
  vol. 32, no. 2, pp. 187--195, 2019.

\bibitem{zabihollahy2020fully}
Fatemeh Zabihollahy, Martin Rajchl, James~A White, and Eranga Ukwatta,
\newblock ``Fully automated segmentation of left ventricular scar from 3d late
  gadolinium enhancement magnetic resonance imaging using a cascaded
  multi-planar u-net (cmpu-net),''
\newblock {\em Medical physics}, vol. 47, no. 4, pp. 1645--1655, 2020.

\bibitem{xu2018mutgan}
Chenchu Xu, Lei Xu, Gary Brahm, Heye Zhang, and Shuo Li,
\newblock ``Mutgan: simultaneous segmentation and quantification of myocardial
  infarction without contrast agents via joint adversarial learning,''
\newblock in {\em International Conference on Medical Image Computing and
  Computer-Assisted Intervention}. Springer, 2018, pp. 525--534.

\bibitem{yang2020simultaneous}
Guang Yang, Jun Chen, Zhifan Gao, Shuo Li, Hao Ni, Elsa Angelini, Tom Wong,
  Raad Mohiaddin, Eva Nyktari, Ricardo Wage, et~al.,
\newblock ``Simultaneous left atrium anatomy and scar segmentations via deep
  learning in multiview information with attention,''
\newblock {\em Future Generation Computer Systems}, vol. 107, pp. 215--228,
  2020.

\bibitem{wong2013influence}
Jorge~A Wong, Raymond Yee, John Stirrat, David Scholl, Andrew~D Krahn, Lorne~J
  Gula, Allan~C Skanes, Peter Leong-Sit, George~J Klein, David McCarty, et~al.,
\newblock ``Influence of pacing site characteristics on response to cardiac
  resynchronization therapy,''
\newblock {\em Circulation: Cardiovascular Imaging}, vol. 6, no. 4, pp.
  542--550, 2013.

\bibitem{sharma2021modified}
Umesh~C Sharma, Kanhao Zhao, Michael Udin, Badri Karthikeyan, and Leslie Ying,
\newblock ``Modified gan augmentation algorithms for the mri-classification of
  myocardial scar tissue in ischemic cardiomyopathy,''
\newblock {\em Frontiers in Cardiovascular Medicine}, p. 1097, 2021.

\bibitem{ridnik2021imagenet21k}
Tal Ridnik, Emanuel Ben-Baruch, Asaf Noy, and Lihi Zelnik-Manor,
\newblock ``Imagenet-21k pretraining for the masses,'' 2021.

\bibitem{ronneberger2015u}
Olaf Ronneberger, Philipp Fischer, and Thomas Brox,
\newblock ``U-net: Convolutional networks for biomedical image segmentation,''
\newblock in {\em International Conference on Medical image computing and
  computer-assisted intervention}. Springer, 2015, pp. 234--241.

\bibitem{chen2021transunet}
Jieneng Chen, Yongyi Lu, Qihang Yu, Xiangde Luo, Ehsan Adeli, Yan Wang, Le~Lu,
  Alan~L. Yuille, and Yuyin Zhou,
\newblock ``Transunet: Transformers make strong encoders for medical image
  segmentation,''
\newblock {\em arXiv preprint arXiv:2102.04306}, 2021.

\bibitem{DBLP:journals/corr/KingmaB14}
Diederik~P. Kingma and Jimmy Ba,
\newblock ``Adam: A method for stochastic optimization,''
\newblock in {\em ICLR (Poster)}, 2015.

\bibitem{wandb}
Lukas Biewald,
\newblock ``Experiment tracking with weights and biases,'' 2020,
\newblock Software available from wandb.com.

\end{thebibliography}

\end{document}